\documentclass[conference, 10pt]{IEEEtran}
\IEEEoverridecommandlockouts
\usepackage{cite}
\usepackage{amsmath,amssymb,amsfonts}
\usepackage{algpseudocode}
\usepackage{graphicx}
\usepackage{textcomp}
\usepackage{xcolor}
\usepackage{braket}
\usepackage{url}
\usepackage{hyperref}
\usepackage{booktabs} 
\usepackage{multirow} 
\usepackage{bm}
\usepackage{physics}

\hypersetup{
  colorlinks   = true, 
  urlcolor     = blue, 
  linkcolor    = blue, 
  citecolor   = blue 
}

\def\BibTeX{{\rm B\kern-.05em{\sc i\kern-.025em b}\kern-.08em
    T\kern-.1667em\lower.7ex\hbox{E}\kern-.125emX}}
\begin{document}

\title{Learning to Learn with Quantum Optimization via Quantum Neural Networks 
\thanks{Corresponding Author: kaun-cheng.chen17@imperial.ac.uk}
}

\author{
\IEEEauthorblockN{
    Kuan-Cheng Chen\IEEEauthorrefmark{2}\IEEEauthorrefmark{3}\IEEEauthorrefmark{1},
    Hiromichi Matsuyama\IEEEauthorrefmark{4},
    Wei-hao Huang \IEEEauthorrefmark{4},
}
\IEEEauthorblockA{\IEEEauthorrefmark{2}Department of Electrical and Electronic Engineering, Imperial College London, London, UK}
\IEEEauthorblockA{\IEEEauthorrefmark{3}Centre for Quantum Engineering, Science and Technology (QuEST), Imperial College London, London, UK}
\IEEEauthorblockA{\IEEEauthorrefmark{4}\textit{Jij Inc.}, 3-3-6 Shibaura, Minato-ku, Tokyo, 108-0023, Japan}
}

\maketitle
\begin{abstract}
Quantum Approximate Optimization Algorithms (QAOA) promise efficient solutions to classically intractable combinatorial optimization problems by harnessing shallow-depth quantum circuits. Yet, their performance and scalability often hinge on effective parameter optimization, which remains nontrivial due to rugged energy landscapes and hardware noise. In this work, we introduce a quantum meta-learning framework that combines quantum neural networks, specifically Quantum Long Short-Term Memory (QLSTM) architectures, with QAOA. By training the QLSTM optimizer on smaller graph instances, our approach rapidly generalizes to larger, more complex problems, substantially reducing the number of iterations required for convergence. Through comprehensive benchmarks on Max-Cut and Sherrington-Kirkpatrick model instances, we demonstrate that QLSTM-based optimizers converge faster and achieve higher approximation ratios compared to classical baselines, thereby offering a robust pathway toward scalable quantum optimization in the NISQ era.
\end{abstract}

\begin{IEEEkeywords}
Quantum Approximate Optimization Algorithm, Quantum Machine Learning, Model Compression, Learning to Learn
\end{IEEEkeywords}

\section{Introduction}
\label{sec:intro}
Optimization problems arise in numerous scientific and engineering contexts, such as logistics\cite{dalal2024digitized}, finance\cite{huber2024exponential}, smart city\cite{mastroianni2023assessing} and communication\cite{gulbahar2025majority,cui2022quantum,chen2025resource}. Although classical optimization algorithms have made considerable progress in addressing these problems\cite{zelinka2012handbook,hidalgo2013comparing,weise2014benchmarking}, they often prove inadequate for large-scale or highly complex tasks due to limited computational resources and intrinsic algorithmic constraints.

Quantum computing has emerged as a transformative computational paradigm that harnesses quantum mechanical effects to tackle problems widely believed to be intractable on classical machines. Within this realm, the Quantum Approximate Optimization Algorithm (QAOA) \cite{farhi2014quantum} has attracted considerable attention for its capacity to solve combinatorial optimization problems, particularly NP-hard ones \cite{chatterjee2024solving}, using comparatively shallow quantum circuits suited to near-term quantum devices \cite{zhou2020quantum,lotshaw2022scaling,chen2024noise}. Despite these advantages, the quest to optimize QAOA parameters remains a formidable challenge due to its non-convex optimization landscape, the presence of numerous local minima, and the practical impact of noise and measurement uncertainties on quantum hardware. In addition, recent work by Bittel \textit{et al.}\ demonstrates that the classical training procedure central to variational quantum algorithms (VQAs) is itself NP-hard; furthermore, under the assumption that \( \bm{P \neq NP} \), no polynomial-time algorithm can achieve an optimization error \( \Delta < 1 \) for all problem instances, highlighting a fundamental limitation in VQA-based hybrid quantum computations\cite{bittel2021training}.

Recent advancements in quantum machine learning (QML) have provided new opportunities to overcome existing challenges in optimization. Quantum machine learning algorithms utilize quantum properties such as entanglement, superposition, and interference to augment classical learning capabilities \cite{biamonte2017quantum,chen2024validating}. Quantum-enhanced recurrent neural networks\cite{li2023quantum,lin2024quantum,hsu2024quantum}, particularly Quantum Long Short-Term Memory (QLSTM) networks \cite{chen2022quantumLSTM}, have emerged as effective tools capable of learning sophisticated optimization strategies by harnessing quantum parallelism and intrinsic quantum correlations\cite{chehimi2024federated,liu2025federated}.

In this work, we propose a novel hybrid quantum-classical framework integrating Quantum Neural Networks (QNNs) with quantum-enhanced learning optimizers to significantly enhance the performance of quantum optimization algorithms. Specifically, we utilize a QLSTM-based optimizer designed to dynamically optimize the parameters of QAOA circuits. This QLSTM-driven optimization scheme enables the learning of generalized strategies that effectively address a broad range of optimization problems. By embedding the QLSTM within a hybrid quantum-classical architecture, our approach realizes a ``learning to learn" paradigm \cite{verdon2019learning}, markedly accelerating convergence and improving the quality of solutions obtained by quantum optimization methods.

This work makes three primary contributions. First, it introduces a hybrid quantum-classical optimization framework that couples QNNs and quantum machine learning optimizers to improve the performance of quantum algorithms. Second, it proposes a QLSTM-based optimizer that learns parameter update rules for QAOA, achieving faster convergence and superior solution quality relative to standard classical methods. Third, it presents an extensive benchmarking procedure against established classical optimizers, including exact solutions to Max-Cut and SK Model instances, thereby establishing a robust reference for evaluating the efficacy of the proposed QLSTM strategy.

The structure of this paper is as follows. Section~\ref{sec:related_works} provides a critical review of recent developments in QAOA, with particular emphasis on hybrid quantum-classical optimization strategies and their inherent limitations. Section~\ref{sec: preliminaries} presents the theoretical foundations of QAOA, establishing a basis for the methodological contributions that follow. We also examine the computational bottlenecks associated with variational parameter training and motivate the need for learning-based optimizers. Section~\ref{sec:meta-learning} details our proposed QLSTM-driven meta-learning framework, including its architectural design, mathematical formulation, and training protocol. Section~\ref{sec:nrd} reports numerical results on both Max-Cut and SK model instances, empirically validating the generalizability, scalability, and performance gains afforded by our method. Finally, Section~\ref{sec:cfw} concludes with a summary of contributions and outlines future research directions, particularly toward broader applications within the landscape of variational quantum algorithms.

\section{Related Works}
\label{sec:related_works}
QAOA has emerged as a promising class of VQAs designed to address a variety of combinatorial optimization problems. Initially proposed by Farhi \textit{et al}.\cite{farhi2014quantum}, QAOA integrates classical optimization techniques with quantum operations to variationally minimize a cost function. Despite its conceptual attractiveness and preliminary empirical successes, the original QAOA formulation frequently encounters challenges such as barren plateaus, slow convergence, and limited approximation ratios at shallow circuit depths. Consequently, extensive research efforts have been dedicated to designing more expressive and efficient ansätze that enhance QAOA performance across diverse problem instances.

A notable research direction involves multi-parameter or multi-angle approaches, which assign distinct parameters to individual qubits or terms in the cost Hamiltonian. For example, ma-QAOA \cite{herrman2022multi} introduces a separate parameter for each element within the cost and mixer Hamiltonians, enabling precise control over local interactions. Similarly, QAOA+ \cite{chalupnik2022augmenting} supplements traditional QAOA layers with an additional, problem-independent yet multi-parameterized layer, improving approximation ratios on random regular graphs. Alternative strategies seek to decrease required circuit depth by incorporating counterdiabatic driving or problem-dependent modifications. Notably, DC-QAOA \cite{chandarana2022digitized,wurtz2022counterdiabaticity} utilizes counterdiabatic terms to enhance convergence rates and approximation quality while maintaining shallow circuit depth.

Incorporating local fields or modifying the mixer Hamiltonian also provides effective avenues for reducing runtime and addressing hardware constraints. For instance, ab-QAOA \cite{yu2022quantum} applies local fields to expedite computation in combinatorial optimization tasks, while ADAPT-QAOA \cite{zhu2022adaptive} iteratively selects mixer Hamiltonians using gradient-based criteria, generating problem-specific circuits dynamically. Recursive QAOA \cite{bravyi2020obstacles} progressively reduces problem size through non-local qubit elimination strategies, overcoming limitations arising from locality constraints.

Broadening the operator space is another compelling approach. For instance, QAOAnsatz \cite{hadfield2019quantum} generalizes conventional cost and mixer operators, facilitating flexible incorporation of both hard and soft constraints. Variants such as GM-QAOA \cite{bartschi2020grover} incorporate Grover-inspired selective phase shifts tailored for vertex covers and traveling salesperson problems, while Th-QAOA \cite{golden2021threshold} replaces the traditional phase separator with a threshold function suitable for Max-Cut and Max Bisection problems. Constraint-preserving mixers \cite{fuchs2022constraint} directly integrate hard constraints within the ansatz, ensuring feasible solutions remain consistently within the solution space.

Another significant research trajectory emphasizes improved initialization and feedback mechanisms. FALQON \cite{magann2022feedback} bypasses classical optimization loops by adjusting parameters based on real-time measurement feedback, achieving monotonically improved approximations with increased circuit depth. Furthermore, FALQON+ \cite{magann2022lyapunov} merges robust initialization from FALQON with QAOA, yielding superior parameter seeds applicable to non-isomorphic graphs.

Finally, a range of advanced techniques has emerged to further enhance QAOA’s performance and introduce problem-specific constraints. For example, FQAOA \cite{yoshioka2023fermionic} employs fermionic particle-number preservation to optimize quantum algorithms in portfolio management. Separately, Quantum Dropout \cite{wang2023quantum} improves algorithmic efficiency by randomly removing circuit components without altering the underlying cost function. Warm-Start QAOA (WS-QAOA) and structure-tuned QAOA (ST-QAOA) \cite{wurtz2021classically} take advantage of approximate classical solutions to construct tailored quantum circuits that can exceed conventional QAOA’s performance at shallow depths. Similarly, modified QAOA \cite{villalba2021improvement} introduces conditional rotations in the cost Hamiltonian to improve approximation ratios for problems such as Max-Cut. Collectively, these developments demonstrate a dynamic and rapidly evolving field, driving QAOA’s scalability, accuracy, and adaptability to more complex problem domains.

In this work, we advance quantum optimization methodologies by introducing a novel framework, ``Learning to Learn with Quantum Optimization via QNN." Our method synergizes QML with VQAs to create an adaptive and expressive quantum-enhanced optimizer capable of learning optimization strategies generalized across diverse problem instances. Significantly, our paper is the first to employ a quantum-enhanced optimizer specifically for tuning QAOA parameters, and we benchmark our approach against Recursive QAOA, the current state-of-the-art method, demonstrating notable performance improvements.

\section{Preliminaries}
\label{sec: preliminaries}

\subsection{Quantum Approximate Optimization Algorithm}

\begin{figure}[!b]
    \centering
    \includegraphics[width=\columnwidth]{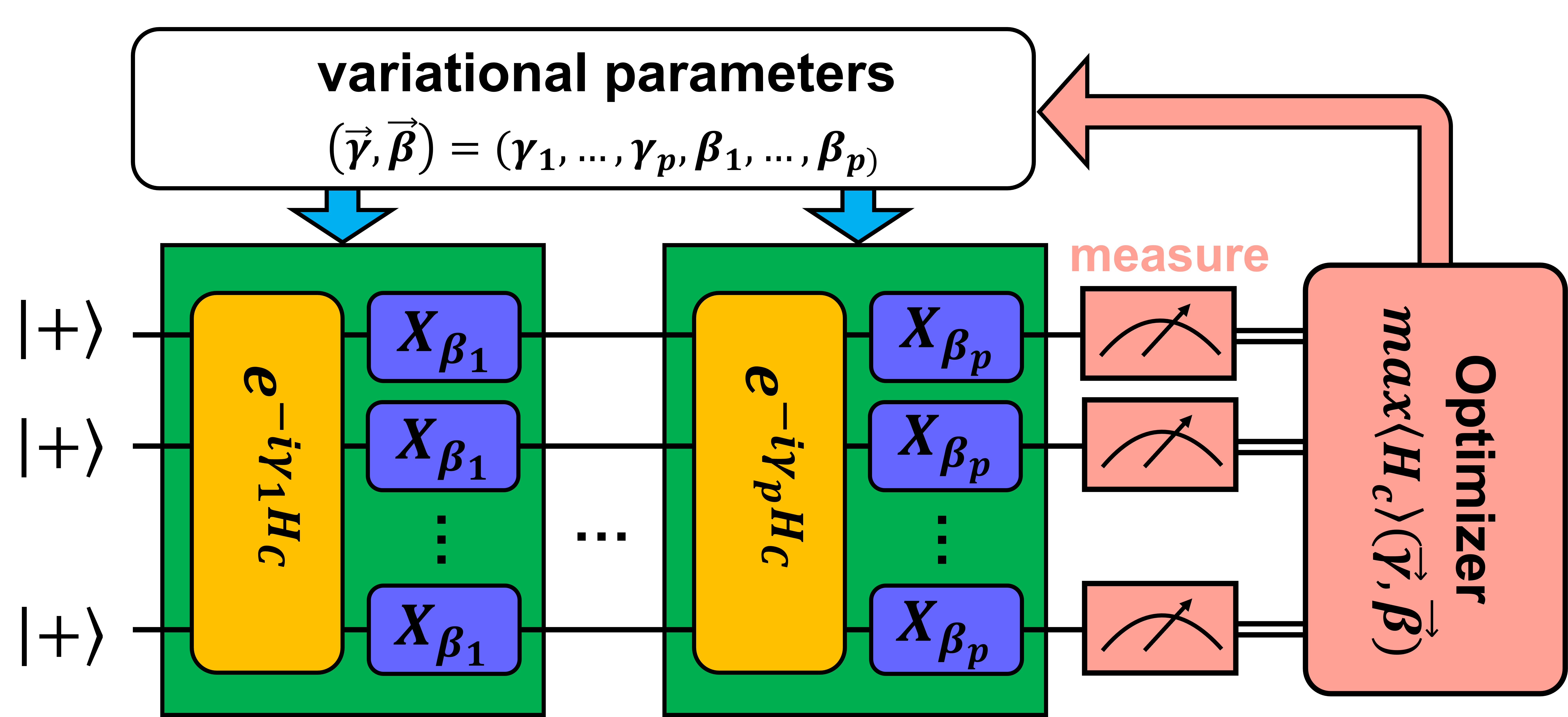}
    \caption{Schematic representation of the QAOA, where a classical optimizer iteratively updates the variational parameters 
$(\vec{\gamma}, \vec{\beta})$ based on measurement outcomes to maximize the expectation value of the cost Hamiltonian 
$\langle H_C \rangle$. Each layer $k$ in the quantum circuit consists of two parts: first, an evolution under the cost Hamiltonian, 
$e^{-i\gamma_k H_C}$, and second, a mixing operation given by 
$e^{-i\beta_k H_B}$, with 
$H_B = \sum_i X_i$. In practice, this mixing evolution can be viewed as a product of single-qubit rotations, 
$\prod_i e^{-i\beta_k X_i}$. After measurement, the resulting bit strings are used to estimate 
$\langle H_C \rangle$, which serves as the objective function for the classical optimizer to update 
$\vec{\gamma}$ and 
$\vec{\beta}$ in subsequent iterations, thereby refining the solution to the optimization problem.
}
    \label{fig:qaoa_architecture}
\end{figure}

The QAOA and its variants~\cite{farhi2014quantum}, which are conceptually inspired by quantum annealing~\cite{hauke2020perspectives} and variational quantum algorithms~\cite{cerezo2021variational}, have been introduced to solve such combinatorial problems on noisy intermediate-scale quantum (NISQ) devices. As depicted in Fig.~\ref{fig:qaoa_architecture}, QAOA aims to approximate the ground state of $H_{C}$ by preparing a variational ansatz of the form
\begin{equation}
\ket{\Phi(\vec{\gamma},\vec{\beta})}
\;=\;
U_{B}(\beta_p)\,U_{C}(\gamma_p)\,\cdots\,
U_{B}(\beta_1)\,U_{C}(\gamma_1)\,\ket{s},
\label{eq:Phi}
\end{equation}
where \( \ket{s} = \ket{+}^{\otimes N} = ((\ket{0}+\ket{1})/\sqrt{2})^{\otimes N} \) is the initial equal superposition state, and \( N \) is the number of qubits. The vectors \( \vec{\gamma} = (\gamma_1, \dots, \gamma_p) \) and \( \vec{\beta} = (\beta_1, \dots, \beta_p) \) denote the real-valued variational parameters. The unitaries $U_{B}$ and $U_{C}$ are defined by
\[
U_{B}(\beta) \;=\; \exp\!\qty(-\mathrm{i}\,\beta\,\sum_{k=1}^{N} X_{k}),
\quad
U_{C}(\vec{\gamma}) \;=\; \exp\!\qty(-\mathrm{i}\,\vec{\gamma}\,H_{C}),
\]
where $X_k$ is the Pauli-$X$ operator acting on the $k$th qubit. One then seeks to minimize the cost function
\begin{equation}
C(\vec{\gamma},\vec{\beta})
\;=\;
\langle \Phi(\vec{\gamma},\vec{\beta})\bigl|H_{C}\bigr|\Phi(\vec{\gamma},\vec{\beta})\rangle
\label{eq:Cgb}
\end{equation}
via a classical optimization routine. The optimal parameters $(\vec{\gamma}, \vec{\beta})$ that maximize $\langle H_C \rangle (\vec{\gamma}, \vec{\beta})$ define an ansatz state $\ket{\Phi(\vec{\gamma}, \vec{\beta})}$ that approximates the ground state of $H_C$. Measuring this state in the computational basis yields the bitstring $\bm{z}^*$ corresponding to an approximate solution of the Max-Cut problem. In principle, increasing the circuit depth $p$ enhances QAOA’s ability to approach the global optimum, though practical limitations of quantum hardware introduce challenges in parameter training and execution.

\subsection{QAOA and its Optimizer}
\label{sec:QAOA-optimizer}

We usually encode optimization problems into the Ising Hamiltonian~$H_C = \sum_{(i,j)\in E} J_{ij}Z_iZ_j$.
Thus, each cost unitary $U_{C}(\gamma_\ell)$ can be regarded as a collection of pairwise interactions $e^{-\mathrm{i}\,\gamma_\ell\,J_{ij}\,Z_i\,Z_j}$ for each edge $(i,j)\in E$, while each mixing unitary $U_{B}(\beta_\ell)$ takes the form $e^{-\mathrm{i}\,\beta_\ell \sum_{k=1}^{N} X_k}$. The number of qubits $N$ and the circuit depth $p$ both directly impact the computational cost of evaluating and optimizing the QAOA parameters, as described below.

A central challenge in QAOA lies in classically optimizing the cost function defined in Eq.~\eqref{eq:Cgb}, where \( \vec{\gamma} = (\gamma_1, \dots, \gamma_p) \) and \( \vec{\beta} = (\beta_1, \dots, \beta_p) \) are real variational parameters. The classical optimizer iteratively proposes parameter updates based on evaluations of \( C(\vec{\gamma}, \vec{\beta}) \) (and possibly its gradient) measured on the quantum device. Suppose one employs a gradient-based method requiring partial derivatives of $C$ with respect to each parameter. A straightforward parameter-shift approach\cite{mitarai2018quantum,wierichs2022general} yields
\begin{align}
\frac{\partial}{\partial \gamma_{\ell}}\,C(\vec{\gamma},\vec{\beta})
&=\, 
\frac{1}{2}
\Bigl[
C\bigl(\ldots,\gamma_{\ell}+\tfrac{\pi}{2},\ldots,\vec{\beta}\bigr) \\
&\quad -
C\bigl(\ldots,\gamma_{\ell}-\tfrac{\pi}{2},\ldots,\vec{\beta}\bigr)
\Bigr],
\nonumber\\[4pt]
\frac{\partial}{\partial \beta_{\ell}}\,C(\vec{\gamma},\vec{\beta})
&=\, 
\frac{1}{2}
\Bigl[
C\bigl(\vec{\gamma},\ldots,\beta_{\ell}+\tfrac{\pi}{2},\ldots\bigr) \\
&\quad -
C\bigl(\vec{\gamma},\ldots,\beta_{\ell}-\tfrac{\pi}{2},\ldots\bigr)
\Bigr].
\label{eq:shift-rule}
\end{align}
Evaluating these expressions involves preparing and measuring $|\Phi(\ldots,\gamma_{\ell}\pm \tfrac{\pi}{2},\ldots,\vec{\beta})\rangle$ or $|\Phi(\vec{\gamma},\ldots,\beta_{\ell}\pm \tfrac{\pi}{2},\ldots)\rangle$ for each parameter. Since there are $2p$ parameters in total, a naive implementation of the parameter-shift rule would require $\mathcal{O}(p)$ circuit executions per iteration to compute all partial derivatives, in addition to the cost of measuring $C$ itself. More sophisticated strategies can sometimes reduce this cost, yet one typically expects the measurement overhead to scale linearly with $p$ when using gradient-based approaches.


The number of gate operations in a QAOA layer fundamentally depends on the number of terms present in the Ising Hamiltonian. Thus, the gate complexity scales as $\mathcal{O}(N^2)$. Additionally, each measurement requires multiple repetitions (often denoted as the number of shots $M$) to accurately estimate the objective function, further increasing the computational cost associated with each parameter update. 



The classical optimization loop typically runs for \( K \) iterations before convergence, where \( K \) depends on both 
the dimension of the parameter space (\( 2p \)) and the optimizer’s hyperparameters (such as learning rate or momentum). 
When the gradient is approximated using parameter-shift rules, the number of quantum circuit evaluations and measurements 
per iteration scales as \( \mathcal{O}(p\,M) \). Multiplying by the total number of iterations \( K \), the overall gate 
count scales as
\begin{equation}
\label{eq:QAOA-complexity}
\mathcal{O}\Bigl(
K\;p\;M\;p_{\mathrm{circuit}}\Bigr),
\quad
p_{\mathrm{circuit}} = \mathcal{O}\bigl(p\,|E|\bigr),
\end{equation}
where \( p_{\mathrm{circuit}} \) characterizes the circuit complexity at each iteration. Consequently, for dense graphs 
with \( |E| \sim \mathcal{O}(N^2) \), each QAOA evaluation step has a circuit depth scaling as 
\(\mathcal{O}(p\,N^2)\), and the total parameter-optimization procedure can grow accordingly in \( p \). 

While the analysis above considers a generic gradient-based optimizer, similar scaling concerns arise for gradient-free 
methods that repeatedly measure \(C(\vec{\gamma},\vec{\beta})\) for various parameter guesses. The next sections will 
demonstrate how an adaptive quantum-classical approach, based on quantum-enhanced recurrent networks, can mitigate the 
complexity of parameter tuning by learning sophisticated parameter-update strategies that reduce the number of 
iterations \(K\) and improve convergence to high-quality Max-Cut solutions.

\section{Meta-Learning with Quantum Long Short-Term Memory Optimizers}
\label{sec:meta-learning}

Optimization of variational parameters in quantum algorithms, such as the QAOA, presents significant challenges due to the high-dimensional and non-convex nature of the parameter landscapes. Traditional optimization techniques often require a substantial number of function evaluations and can become computationally prohibitive as the problem size scales. To address these challenges, we introduce a meta-learning framework that leverages QLSTM networks to serve as neural optimizers for QAOA parameter tuning. This approach harnesses the capabilities of recurrent neural networks (RNNs) within a quantum-enhanced architecture to learn adaptive optimization strategies, effectively enabling the system to "learn to learn."

\subsection{Meta-Learning Framework for Quantum Neural Networks}

Meta-learning, frequently referred to as “learning to learn” \cite{finn2019online,tian2022meta}, seeks to train models that leverage prior experiences from multiple related tasks to enhance their learning efficiency. In the domain of quantum optimization, this approach focuses on creating optimizers capable of generalizing across various QNN instances, thereby reducing the computational overhead required for parameter tuning. Here, we adopt the structure of a classical LSTM—retaining its gating and memory mechanisms—but implement it as a quantum neural network, enabling dynamic updates of circuit parameters to tackle target problem classes such as Max-Cut.

Formally, consider a family of variational quantum circuits described by parameters \((\vec{\gamma}, \vec{\beta})\), for which the goal is to minimize the cost function
\begin{equation}
f(\vec{\gamma}, \vec{\beta}) = \langle \psi(\vec{\gamma}, \vec{\beta}) \,|\, \hat{H} \,|\, \psi(\vec{\gamma}, \vec{\beta}) \rangle,
\end{equation}
where \(|\psi(\vec{\gamma}, \vec{\beta})\rangle\) denotes the quantum state prepared by the QAOA circuit, and \(\hat{H}\) is the problem Hamiltonian. In this paper, we focus on QAOA, where \(\hat{H}\) encodes the Max-Cut objective (as introduced in Eq.~\eqref{eq:Cgb}). Our meta-learning framework employs a QLSTM optimizer to recommend parameter updates for \((\vec{\gamma}, \vec{\beta})\) based on historical observations of cost values and parameter adjustments. While we demonstrate this approach specifically for QAOA on Max-Cut, the same QLSTM-based strategy is broadly applicable to other Variational Quantum Algorithms, simply by substituting the relevant cost Hamiltonian.

\subsection{Long Short-Term Memory Networks}
\begin{figure}[!b]
\centering
\includegraphics[width=\columnwidth]{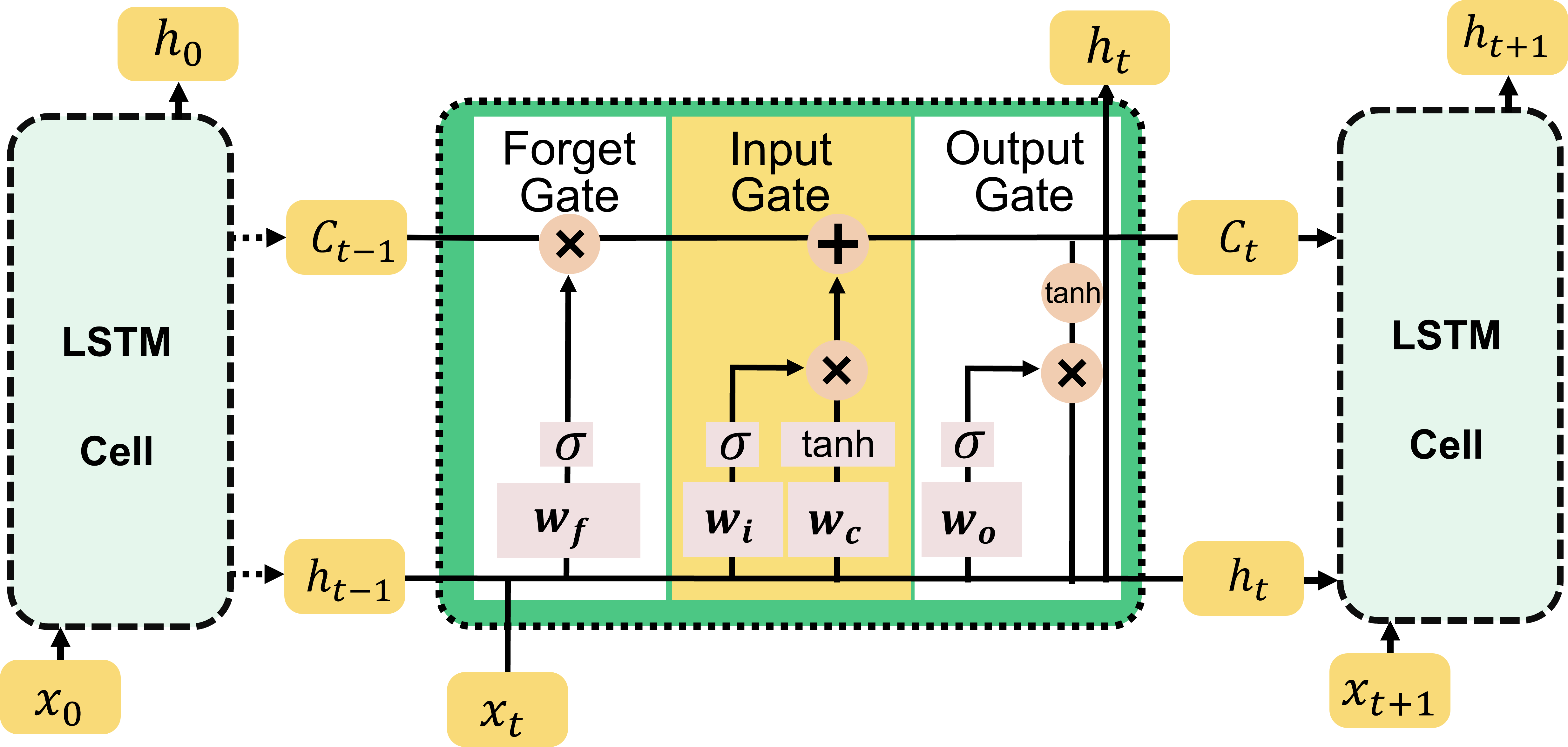}
\caption{Schematic of a standard LSTM cell. The input gate, forget gate, and output gate regulate the information flow into, through, and out of the cell state ($C_t$). This gating mechanism enables the network to capture both short-term and long-term dependencies in the data.}
\label{fig:lstm_architecture}
\end{figure}

Long Short-Term Memory (LSTM) networks \cite{LSTM} are a class of Recurrent Neural Networks (RNNs) \cite{RNN} designed to overcome the vanishing and exploding gradient issues frequently observed in basic RNNs. As illustrated in Fig.~\ref{fig:lstm_architecture}, the LSTM cell employs a gating mechanism—comprising the input, forget, and output gates—to manage the flow of information across time steps. This structure enables the selective retention of long-term dependencies while preserving essential short-term information, thereby ensuring more stable gradient propagation compared to conventional RNNs.

Beyond their success in sequence modeling tasks such as language processing and time series forecasting \cite{chen2016enhanced,yao2018improved,alkin2024vision,kilinc2024multimodal}, LSTMs also hold significant promise for meta-learning applications. Meta-learning, sometimes referred to as “learning to learn,” focuses on training an optimizer or model to generalize across multiple tasks by leveraging knowledge from prior learning experiences. In the context of quantum algorithms such as QAOA, the optimization landscape is often noisy and requires iterative adjustments of a parameterized quantum circuit. LSTMs excel in this setting for several reasons:

\begin{itemize}
\item \textbf{Sequential Parameter Updates:} QAOA parameters are updated iteratively, forming a natural temporal sequence. LSTMs are well-suited to handle sequential data due to their recurrent design, allowing them to track information across multiple optimization steps.
\item \textbf{Adaptive Memory Mechanism:} Through their gating mechanism, LSTMs learn to store or discard information based on its long-term relevance. This capacity is essential when optimizing QAOA parameters over successive iterations, as the optimizer must balance short-term fluctuations (e.g., noise in cost function evaluations) with overarching trends.

\item \textbf{Resistance to Vanishing/Exploding Gradients:} The cell state and gating operations in LSTMs mitigate common gradient flow problems in deep unrolled computational graphs. Because QAOA-based training typically involves multiple update steps—each requiring backpropagation through time—this stability is crucial for robust convergence.

\item \textbf{Black-Box Optimization Compatibility:} In many realistic quantum settings, exact gradients of the cost function may be difficult to obtain or inherently noisy. LSTMs can learn parameter update heuristics based on sequence-to-sequence modeling, allowing them to function even when explicit gradient information is limited.

\end{itemize}

These properties make LSTMs a natural fit for meta-learning optimizers in quantum variational algorithms. By unrolling an LSTM over the sequence of QAOA updates, the network can internalize strategies that adapt to noise, partial gradient information, and other complexities encountered in near-term quantum devices. The next section explores a quantum-enhanced variant of this approach, QLSTM, tailored specifically for parameter optimization in QAOA.

\subsection{Quantum Long Short-Term Memory Networks}

LSTM networks are a type of RNN capable of learning long-term dependencies in sequential data \cite{hochreiter1997long}. Extending this concept to the quantum domain, QLSTM architecture has been proposed by Chen \textit{et al}.\cite{chen2022quantumLSTM}, which integrates variational quantum circuits within the LSTM cell to process quantum-specific information effectively.

The QLSTM cell comprises three primary components: the forget gate, the input and update gates, and the output gate, each implemented using VQCs. Mathematically, the operations within a QLSTM cell can be described as:
\begin{subequations}
    \label{eq:qlstm}
    \begin{align}
        f_t &= \sigma\left(\text{VQC}_1(v_t)\right), \label{eq:qlstm_f} \\
        i_t &= \sigma\left(\text{VQC}_2(v_t)\right), \label{eq:qlstm_i} \\
        \tilde{C}_t &= \tanh\left(\text{VQC}_3(v_t)\right), \label{eq:qlstm_Ctilde} \\
        C_t &= f_t \odot C_{t-1} + i_t \odot \tilde{C}_t, \label{eq:qlstm_C} \\
        o_t &= \sigma\left(\text{VQC}_4(v_t)\right), \label{eq:qlstm_o} \\
        h_t &= \text{VQC}_5\left(o_t \odot \tanh(C_t)\right), \label{eq:qlstm_h} \\
        y_t &= \text{VQC}_6\left(o_t \odot \tanh(C_t)\right), \label{eq:qlstm_y}
    \end{align}
\end{subequations}
where $v_t = [h_{t-1}, \theta_t]$ is the concatenated input comprising the previous hidden state $h_{t-1}$ and the current QNN parameters $\theta_t$. The symbol $\sigma$ denotes the sigmoid activation function, and $\odot$ represents element-wise multiplication. Each VQC is parameterized by its own set of variational parameters and is responsible for processing the input to generate gate activations and transformations within the QLSTM cell.

\begin{figure}[t]
    \centering
    \includegraphics[width=\columnwidth]{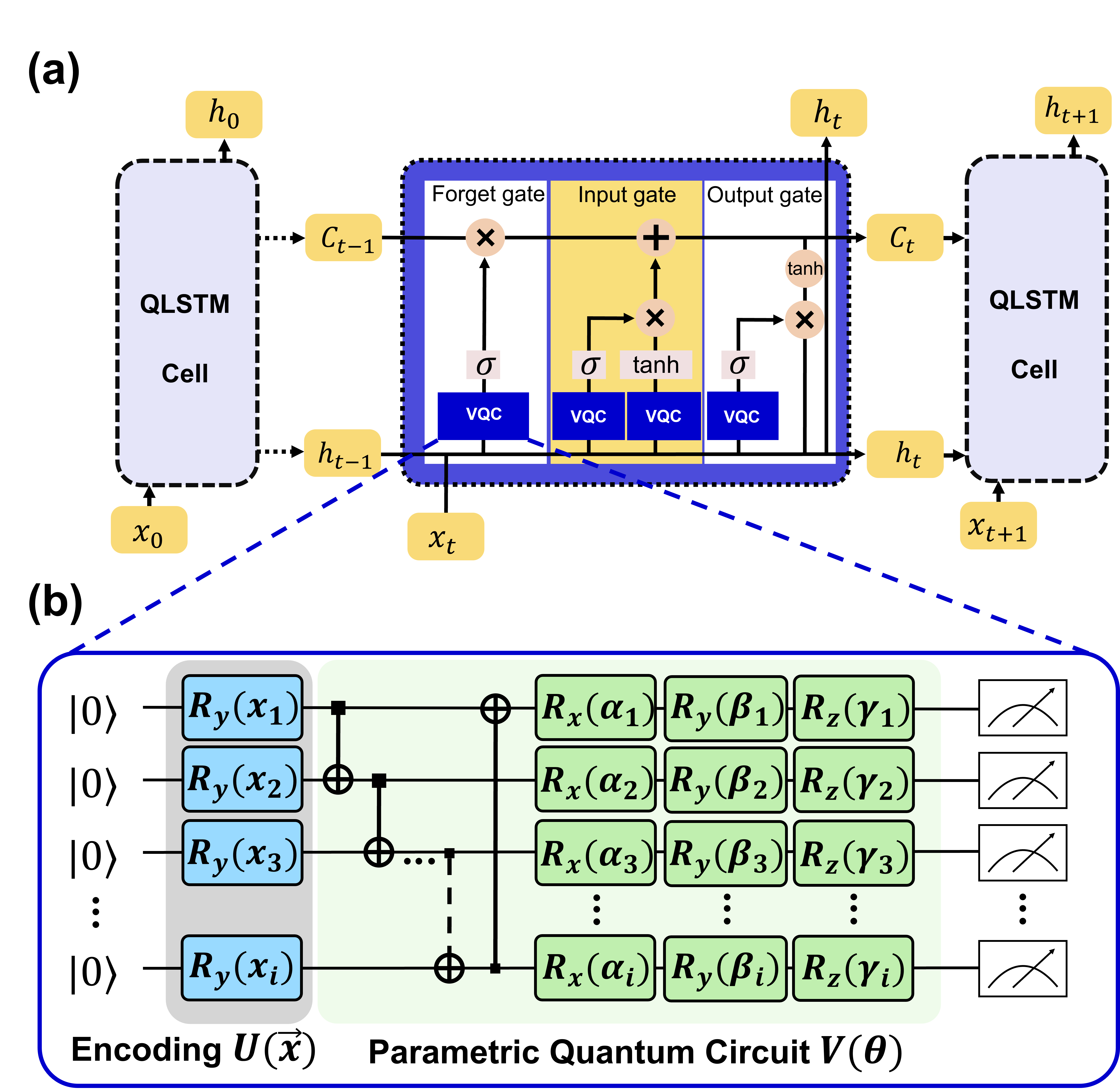}
    \caption{Schematic diagram of the QLSTM architecture.  
(a) Overview of the QLSTM cell, where classical nonlinear operations in standard LSTM gates are replaced with variational quantum circuits (VQCs), enabling quantum-enhanced memory processing.  
(b) Detailed structure of the VQC used within each QLSTM gate. The circuit comprises a data encoding layer \(U(\vec{x})\), where classical inputs \(\vec{x} = (x_1, x_2, \dots, x_i)\) are mapped to quantum states via parameterized rotations (e.g., \(R_y(x_i)\)), followed by a trainable variational block \(V(\vec{\theta})\), consisting of parameterized single-qubit gates (e.g., \(R_x, R_y, R_z\)) and entangling operations. Measurement outcomes provide nonlinear transformations that act as gate activations within the QLSTM cell.}
    \label{fig:qlstm_architecture}
\end{figure}

Figure~\ref{fig:qlstm_architecture} schematically illustrates the QLSTM architecture, highlighting the integration of VQCs within the LSTM cell structure.

\subsection{QLSTM as a Meta-Learning Optimizer for QAOA}

To utilize the QLSTM as a meta-learning optimizer for QAOA, we adopt the following procedure:

1. \textbf{Parameter Update Rule:} At each optimization step $t$, the QLSTM receives the current QAOA parameters $\theta_t$ and the corresponding cost function evaluation $y_t = f(\theta_t)$. The QLSTM processes this information to generate an updated parameter set $\theta_{t+1}$:
\begin{equation}
\theta_{t+1} = \text{QLSTM}_\phi\left(h_t, \theta_t, y_t\right),
\label{eq:parameter_update}
\end{equation}
where $\phi$ represents the trainable parameters of the QLSTM network.

\begin{figure}[!b]
    \centering
    \includegraphics[width=\columnwidth]{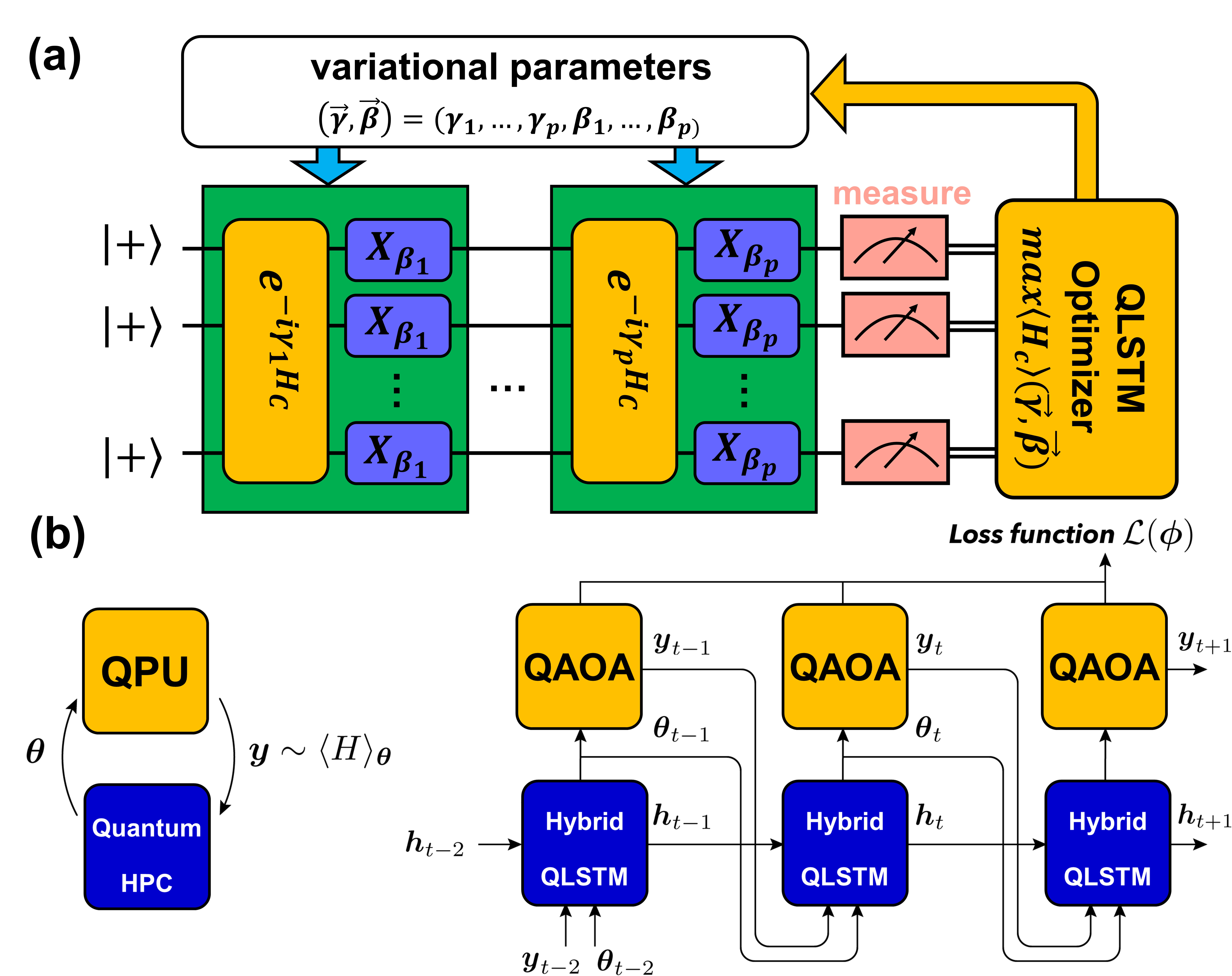}
    \caption{Schematic representation of the QAOA framework enhanced with a QLSTM optimizer.
(a) The parameterized quantum circuit follows the standard QAOA structure but replaces the classical optimizer with a recurrent quantum-classical model. Following measurement, the estimated expectation value \(\langle H_C \rangle\) is provided to a QLSTM-based optimizer, which learns to generate optimized variational parameters \((\vec{\gamma}, \vec{\beta})\) by capturing temporal dependencies across optimization steps. 
(b) Architecture of the QLSTM optimizer. At each time step \(t\), a hybrid QLSTM block receives the current and past QAOA outputs, including previous parameters \(\theta_{t-1}\) and measurement outcomes \(y_{t-1}\), along with internal hidden states \(h_{t-1}\). The hybrid model integrates classical processing with quantum subroutines on a QPU to generate the next parameter set \(\theta_t = (\vec{\gamma}_t, \vec{\beta}_t)\), enabling adaptive and history-aware learning dynamics that aim to improve convergence and generalization in variational quantum algorithms.}
    \label{fig:qlstm_optimizer_architecture}
\end{figure}

2. \textbf{Training the QLSTM:} The QLSTM is trained across multiple QAOA instances to learn an effective parameter update strategy. The training objective is to minimize a meta-loss function that captures the optimization performance over a distribution of QAOA tasks. One suitable choice of meta-loss is the cumulative improvement in the cost function:
\begin{equation}
\mathcal{L}(\phi) = \mathbb{E}\left[\sum_{t=1}^{T} \max\left(0, y_t^{\text{best}} - y_t\right)\right],
\label{eq:meta_loss}
\end{equation}
where $y_t^{\text{best}} = \min_{j < t} y_j$ denotes the best cost function value observed up to step $t$. This loss function encourages the QLSTM to propose parameter updates that lead to continual improvement in the optimization process.

3. \textbf{Gradient-Based Optimization:} Training the QLSTM involves backpropagating gradients through the unrolled optimization process. Given the recurrent nature of the QLSTM, this is achieved using backpropagation through time (BPTT), which updates the QLSTM parameters $\phi$ to minimize the meta-loss $\mathcal{L}(\phi)$. Due to the complexities associated with quantum circuits and their inherent noise, the training is performed using simulated quantum environments where gradients can be accurately estimated.

4. \textbf{Integration with QAOA}: Once trained, the QLSTM serves as an optimizer for QAOA by providing intelligent parameter updates based on the historical performance of parameter sets.

\subsection{Mathematical Framework and Derivations}

To formalize the interaction between QLSTM and QAOA, consider the optimization trajectory $\{\theta_t\}_{t=1}^T$ generated by the QAOA algorithm under the guidance of the QLSTM optimizer. The QLSTM aims to minimize the cumulative meta-loss over this trajectory, defined as:
\begin{equation}
\mathcal{L}(\phi) = \mathbb{E}_{\mathcal{T}}\left[\sum_{t=1}^{T} \max\left(0, f(\theta_t^{\text{best}}) - f(\theta_t)\right)\right],
\label{eq:trajectory_loss}
\end{equation}
where $\mathcal{T}$ represents the distribution of QAOA instances (e.g., different Max-Cut problem graphs), and $\theta_t^{\text{best}} = \arg\min_{j \leq t} f(\theta_j)$ is the best parameter set up to step $t$.

The gradient of the meta-loss with respect to the QLSTM parameters $\phi$ is given by:
\begin{equation}
\nabla_\phi \mathcal{L}(\phi) = \mathbb{E}_{\mathcal{T}}\left[\sum_{t=1}^{T} \nabla_\phi \max\left(0, f(\theta_t^{\text{best}}) - f(\theta_t)\right)\right].
\label{eq:gradient_meta_loss}
\end{equation}
This gradient is computed using BPTT, which propagates the error signals from the meta-loss back through each time step of the QLSTM and the corresponding QAOA evaluations.

\begin{figure}[!b]
    \centering
    \includegraphics[width=\columnwidth]{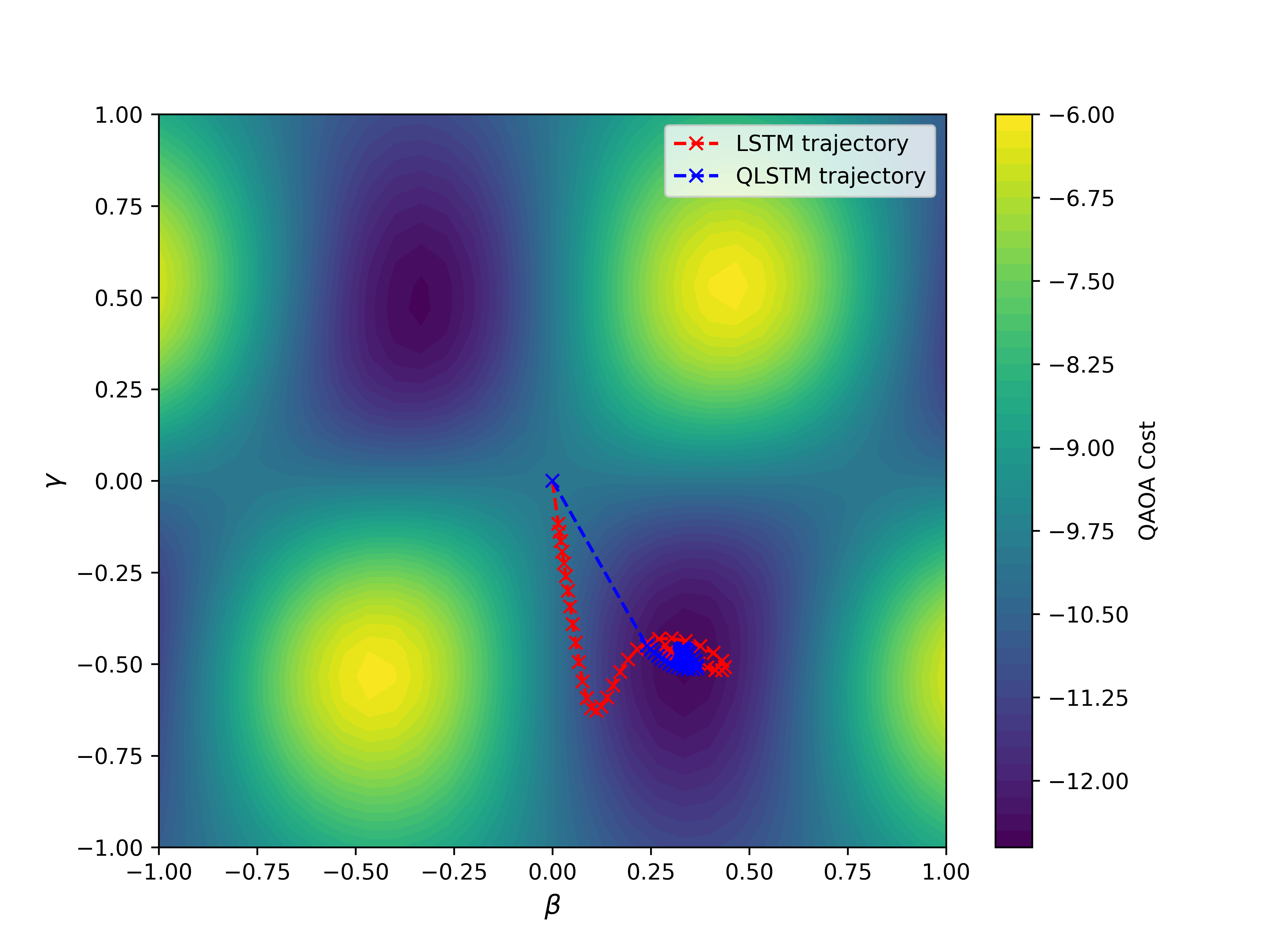}
    \caption{Comparison of LSTM (red) and QLSTM (blue) optimization trajectories in the single-layer QAOA parameter space, evaluated on a randomly generated graph with 10 nodes and edge probability \( P = 0.4 \). The figure highlights the faster convergence and lower final cost achieved by the QLSTM optimizer during training. Both trajectories start from the same initial point (Step 0 at \( \gamma = 0, \beta = 0 \)).}
    \label{fig:qlstm_trajectory}
\end{figure}

The QLSTM's ability to capture temporal dependencies and historical optimization performance enables it to propose parameter updates that are informed by the optimization trajectory (see Fig.~\ref{fig:qlstm_trajectory}), thereby enhancing the efficiency and effectiveness of the QAOA optimization process.

From a resource perspective, for a QAOA of depth $p$ acting on an $n$-qubit system, the quantum circuit requires $n$ qubits and optimizes over $2p$ variational parameters corresponding to alternating applications of cost and mixer unitaries. In contrast, the QLSTM-based optimizer leverages parameterized VQCs that use a fixed number of qubits per cell, specifically $1 + 4p$ qubits, where $\text{input \. size} = 1 + 2p$ and $\text{hidden \. size} = 2p$. Consequently, while the QAOA problem Hamiltonian scales with $n$, the quantum resource requirements of the QLSTM optimizer are independent of $n$ and depend only on the circuit depth $p$. This separation makes the QLSTM-based meta-optimization framework highly scalable and parameter-efficient for large problem instances.

\subsection{Advantages and Complexity Reduction}
\label{subsec:transfer_learning_qaoa}

The integration of QLSTM-based meta-learning optimizers with QAOA offers multiple practical benefits, particularly in terms of scalability and computational efficiency. First, by learning sophisticated parameter-update heuristics, the QLSTM greatly reduces the number of optimization iterations $K$ required to reach high-quality solutions, thereby mitigating the overall computational cost prescribed by Eq.~\eqref{eq:QAOA-complexity}. Second, the QLSTM’s ability to learn a problem-agnostic update strategy facilitates transfer learning: once trained on smaller quantum optimization problems, it can be directly applied to larger instances with only moderate fine-tuning. Third, the recurrent structure of the QLSTM, equipped with internal memory states, enables it to navigate non-convex parameter landscapes more effectively than conventional methods, thus reducing the risk of stagnation in local minima.

To formalize the computational complexity for a QLSTM-optimized QAOA, let \( T_{\text{meta}} \) be the unrolled time horizon for the meta-optimization routine, and \( K_{\text{meta}} \) the number of meta-training iterations. Denoting by \( M \) the number of shots for estimating expectation values, by \( \lvert E \rvert \) the number of edges for a given Max-Cut instance, and by \( p \) the QAOA circuit depth, one can express the total runtime scaling as

\begin{equation}
\label{eq:meta_complexity_repeat}
\mathcal{O}(K_{\text{meta}}\,T_{\text{meta}}\,p\,\lvert E \rvert\,M).
\end{equation}

This includes both the forward execution of QAOA and the offline training overhead for the QLSTM. Although training the QLSTM incurs an upfront cost, its learned update rules can reduce the effective iteration count \( K \) required for convergence, potentially offering a net improvement over classical optimizers. Moreover, once trained, the QLSTM-based approach generalizes to various circuit depths and problem instances, making it a versatile tool for a wide range of variational quantum algorithms.

\subsubsection{Transfer Learning from Smaller to Larger QAOA Instances}

A central feature of the QLSTM-based meta-optimizer is its capacity to generalize from smaller problem instances to larger ones. Formally, let $\{\theta_t\}_{t=1}^T$ denote the QAOA parameter trajectories generated on $N_1$-qubit (or $N_1$-node) graphs. Suppose we gather sufficient training data to approximate the distribution of cost landscapes for problems of size $N_1$. We then train the QLSTM parameters $\phi$ to minimize a meta-loss of the form (same as Eq.~\eqref{eq:trajectory_loss})
\begin{equation}
\mathcal{L}(\phi) \;=\; \mathbb{E}_{\mathcal{T}_{N_1}}\!\Bigl[
\sum_{t=1}^{T} 
\max\bigl(0,\,f(\theta_t^{\mathrm{best}}) - f(\theta_t)\bigr)
\Bigr],
\label{eq:meta_loss_smaller}
\end{equation}
where $\mathcal{T}_{N_1}$ indicates the distribution of QAOA instances with $N_1$ vertices (or qubits).

Once the QLSTM has converged on these smaller problems, we apply the learned meta-optimizer to larger instances with $N_2 > N_1$. Concretely, we initialize a QAOA problem of size $N_2$ with parameters $\{\theta'_t\}$ and update them using the QLSTM policy:
\begin{equation}
\theta'_{t+1} \;=\; \mathrm{QLSTM}_{\phi}\bigl(h'_t,\theta'_t, y'_t\bigr),
\label{eq:meta_update_larger}
\end{equation}
where $y'_t = f(\theta'_t)$ is the cost function value on the larger instance, and $h'_t$ is the recurrent hidden state propagated through the unrolled QLSTM. Notably, the same QLSTM parameters $\phi$ are retained, even though $N_2 \neq N_1$. This transfer learning paradigm leverages the insight that the QAOA parameter landscapes for different system sizes often share structural similarities, enabling a meta-optimizer trained on $N_1$-node graphs to propose effective search directions for $N_2$-node graphs.

We have verified this transfer learning capability by training the QLSTM optimizer on randomly generated Max-Cut instances with $N_1 = 7$ nodes and then evaluating its performance on larger instances with $8$--$16$ nodes. Empirically, we observed that the QLSTM-trained policy substantially accelerated convergence compared to baseline optimizers such as Stochastic Gradient Descent, Adam, and Nelder-Mead, without requiring significant modifications to the underlying QLSTM. Mathematically, if $\hat{T}_{\text{large}}$ is the average number of iterations needed for larger instances under the QLSTM policy, and $T_{\text{baseline}}$ the iterations for the best-performing classical optimizer, we have
\begin{equation}
\hat{T}_{\text{large}} \;\ll\; T_{\text{baseline}},
\end{equation}
showcasing a nontrivial reduction in iteration count. Consequently, the total runtime for large-scale instances also decreased, as predicted by the meta-learning complexity analysis in Eq.~\eqref{eq:meta_complexity_repeat}.


From a quantum-theoretical perspective, although increasing $N$ inevitably modifies the individual weights $W_{ij}$ and thus reshapes the global cost landscape, the \textit{local structure} of the QAOA cost Hamiltonian remains fundamentally unchanged. In particular, 
$H_C = \sum_{(i,j)\in E} W_{ij} \, Z_i \, Z_j$
is invariably expressed as a sum of pairwise $Z_i Z_j$ interactions, preserving its operator form despite varying qubit counts or graph topologies. Consequently, the principle of alternating cost and mixing unitaries endures, and architectures such as QLSTMs can detect and leverage these consistent local features. Hence, even as $N$ grows and the Hilbert space dimension expands, a suitably trained meta-optimizer can identify universal characteristics of the parameter landscape that remain predictive across diverse instances, thereby facilitating effective transfer learning. 


\section{Numerical Results and Discussion}
\label{sec:nrd}

By training on small QAOA instances and deploying the QLSTM optimizer on larger graphs, one achieves a remarkable balance between computational expediency and optimization accuracy. The scaling advantages become increasingly pronounced for dense graphs or deep QAOA circuits. Consequently, the QLSTM meta-learning optimizer not only provides theoretical benefits—such as reduced iteration counts and greater resilience to noise—but also demonstrates tangible improvements in practical, larger-scale quantum optimization tasks.

\subsection{Max-Cut Problems}

\begin{figure*}[htpb]
    \centering
    \includegraphics[width=2.1\columnwidth]{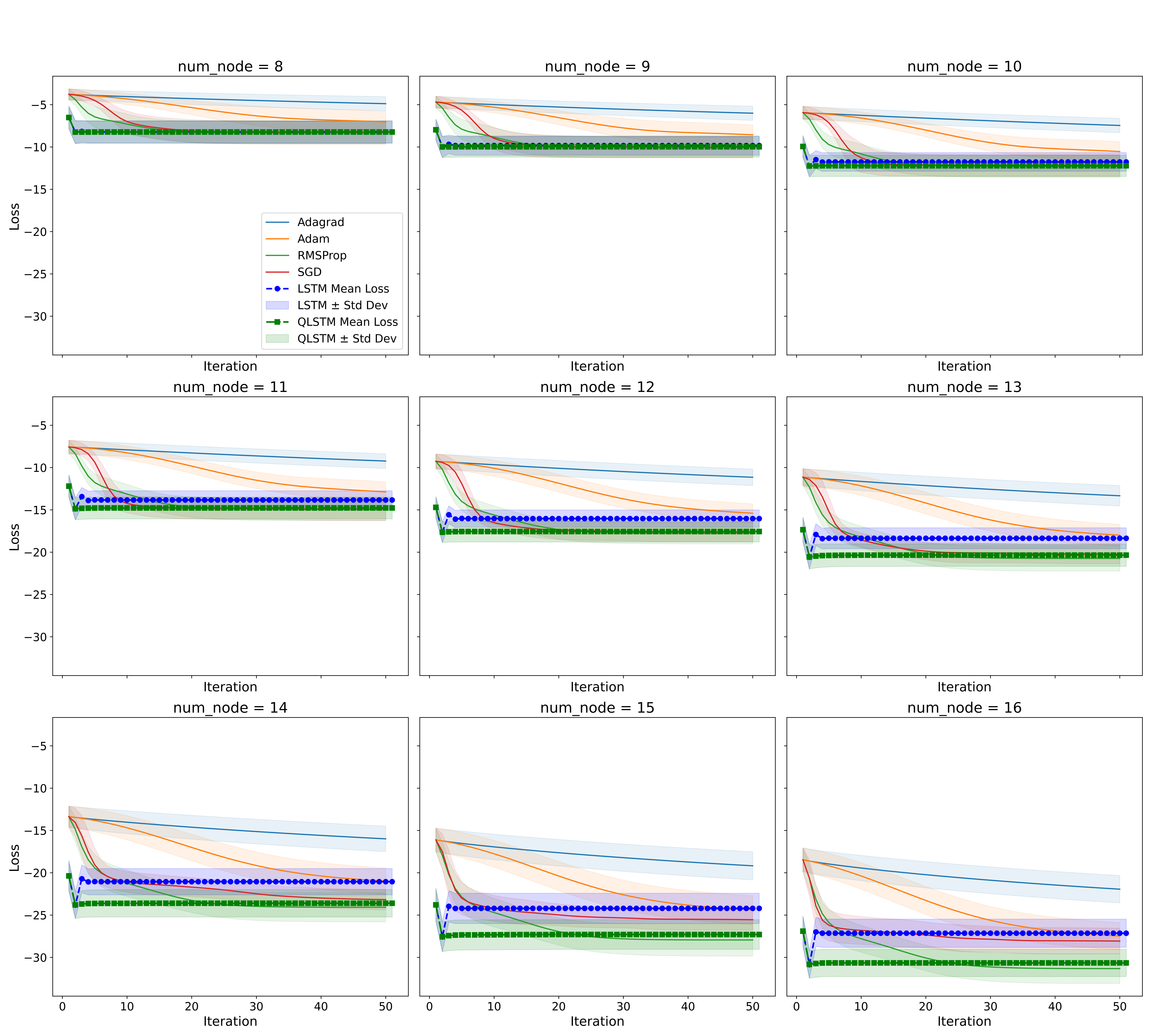}
    \caption{Loss trajectories over 50 inference iterations for various classical and learning-based optimizers, evaluated on Max-Cut instances with \(P=3/7\) (as demonstration) and varying numbers of nodes (8 to 16). The QLSTM-QAOA framework (green) demonstrates superior convergence behavior compared to classical optimizers (Adagrad, Adam, RMSProp, and SGD), achieving consistently lower loss values with minimal variance across different problem sizes. LSTM (blue) also exhibits robust performance but with slightly higher fluctuations.}
    \label{fig:qlstm_qaoa_benchmark}
\end{figure*}

\begin{table*}[!t]
\centering
\caption{Approximation Ratios at Iteration=3 (Mean $\pm$ Std) for $p$-values $2/7$ through $6/7$.}
\label{tab:approx_ratio_iter2}
\begin{tabular}{lcccccc|c}
\toprule
\textbf{$P$-value} & \textbf{QLSTM} & \textbf{LSTM} & \textbf{RMSProp} & \textbf{SGD} & \textbf{Adam} & \textbf{Adagrad} & \textbf{R-QAOA} \cite{finvzgar2024quantum} \\
\midrule
\textbf{2/7} & \textbf{0.82 $\pm$ 0.12} & 0.75 $\pm$ 0.10 & 0.54 $\pm$ 0.09 & 0.45 $\pm$ 0.09 & 0.41 $\pm$ 0.08 & 0.41 $\pm$ 0.08 & 0.92 $\pm$ 0.04 \\
\textbf{3/7} & \textbf{0.87 $\pm$ 0.0}9 & 0.79 $\pm$ 0.07 & 0.59 $\pm$ 0.08 & 0.52 $\pm$ 0.12 & 0.47 $\pm$ 0.08 & 0.46 $\pm$ 0.08 & 0.90 $\pm$ 0.05 \\
\textbf{4/7} &\textbf{ 0.90 $\pm$ 0.07} & 0.84 $\pm$ 0.07 & 0.65 $\pm$ 0.09 & 0.62 $\pm$ 0.17 & 0.52 $\pm$ 0.08 & 0.51 $\pm$ 0.08 & 0.89 $\pm$ 0.07 \\
\textbf{5/7} & \textbf{0.92 $\pm$ 0.05} & 0.89 $\pm$ 0.06 & 0.72 $\pm$ 0.09 & 0.72 $\pm$ 0.18 & 0.56 $\pm$ 0.09 & 0.56 $\pm$ 0.09 & 0.87 $\pm$ 0.06 \\
\textbf{6/7} & \textbf{0.93 $\pm$ 0.03} & 0.93 $\pm$ 0.05 & 0.75 $\pm$ 0.08 & 0.80 $\pm$ 0.16 & 0.57 $\pm$ 0.08 & 0.56 $\pm$ 0.08 & 0.86 $\pm$ 0.07 \\
\bottomrule
\end{tabular}
\end{table*}

\textbf{Definition (Max-Cut problem):\\} 

\begin{figure}[!t]
    \centering
    \includegraphics[width=\columnwidth]{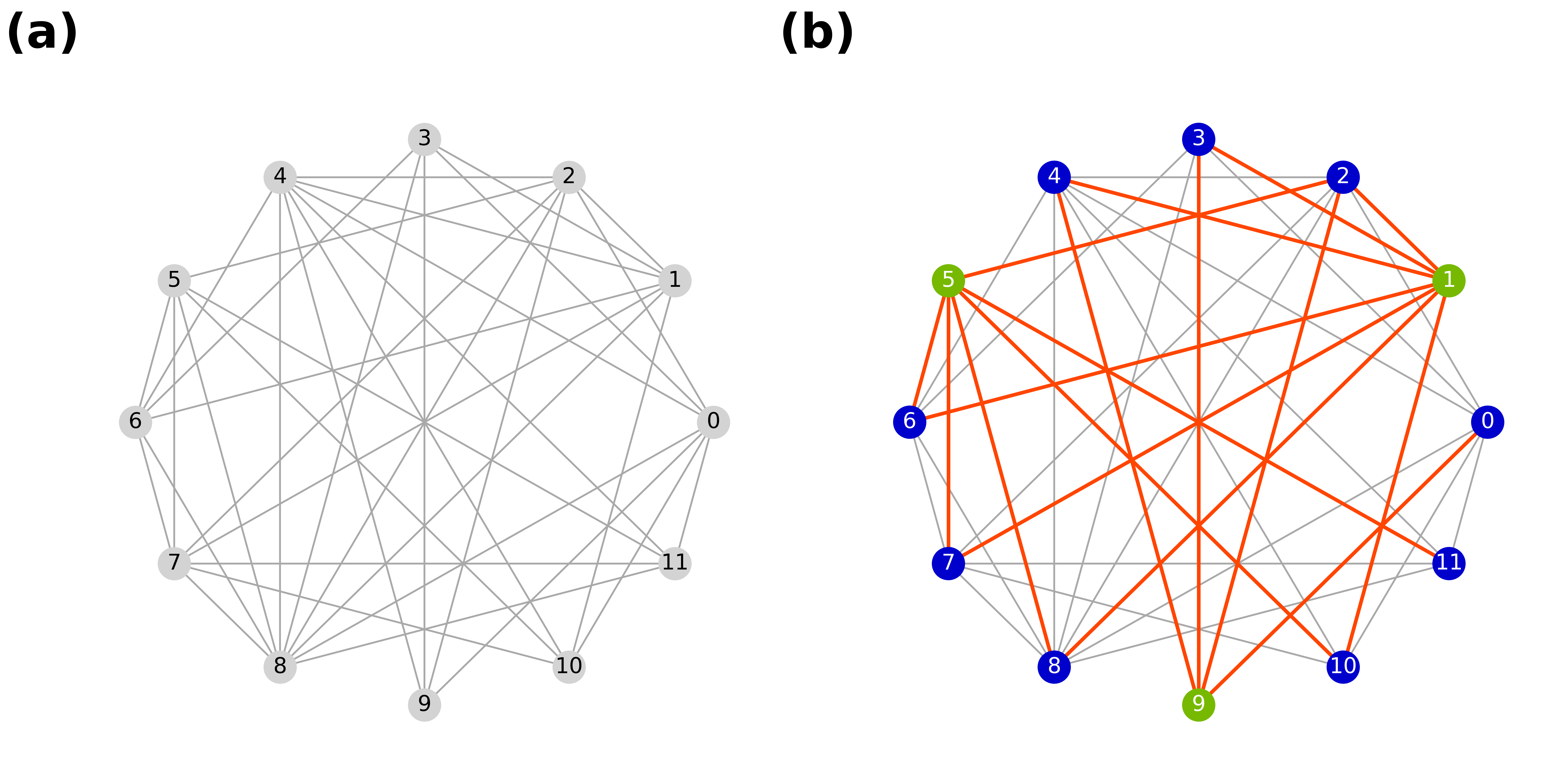}
    \caption{Max-Cut Problem and Solution on a 12-Node Graph. (a) Illustration of the Max-Cut problem instance with 60\% edge connectivity. (b) Solution showing the partition of nodes into two groups, with cut edges highlighted in orange red.}
    \label{Max-Cut}
\end{figure}

Let \(G = (V, E)\) be a graph with edge weights \(w_{ij} \in \mathbb{R}\) for \((i,j) \in E\). In the \(\{+1, -1\}\) representation, each vertex \(i \in V\) is assigned a binary variable \(z_i \in \{+1, -1\}\). A cut is therefore described by the bitstring \(\bm{z} \in \{+1, -1\}^{|V|}\), where \(z_i = +1\) indicates that vertex \(i\) belongs to one subset, and \(z_i = -1\) indicates it belongs to the other. The Max-Cut cost function can then be written as
\begin{equation}
C(\bm{z}) 
\;=\; \frac{1}{2} \sum_{(i,j)\in E} w_{ij}\,\bigl(1 - z_i\,z_j\bigr) 
\;=\; c \;-\; \frac{1}{2} \sum_{(i,j)\in E} w_{ij}\,z_i\,z_j,
\label{eq:Max-Cut-def}
\end{equation}
where
\begin{equation}
c \;=\; \frac{1}{2} \sum_{(i,j)\in E} w_{ij}
\end{equation}
is a constant offset that depends only on the problem instance. Finding the bitstring \(\bm{z}^*\) that maximizes \(C(\bm{z})\) is known to be NP-hard~\cite{arora1998approximability}. Consequently, one often resorts to approximation algorithms in practice. A notable example is the Goemans–Williamson algorithm~\cite{karloff1996good}, which uses semidefinite programming and achieves an approximation ratio of at least \(0.878\)~\cite{goemans1995improved}.

To represent the Max-Cut problem within a quantum framework, we define the cost Hamiltonian as
\begin{equation}
H_{C} 
\;=\;
\sum_{(i,j)\in E} w_{ij}\,Z_i\,Z_j,
\label{eq:HC}
\end{equation}
where \(Z_i\) is the Pauli-\(Z\) operator acting on qubit \(i\). This Hamiltonian is diagonal in the computational basis, and its ground state corresponds to the bitstring that maximizes the total cut weight.

In the context of the Max-Cut problem, the \emph{approximation ratio} for a candidate solution $\bm{z}$ is defined as the ratio of its achieved cut value to that of the optimal cut value:
\begin{equation*}
  \text{Approx.~Ratio}(\bm{z}) \;=\; \frac{C(\bm{z})}{C(\bm{z}^*)},
\end{equation*}
where $C(\bm{z}^*)$ is the maximum cut cost of the given graph [see Eq.~\eqref{eq:Max-Cut-def}]. By construction, $0 \leq \text{Approx.~Ratio}(\bm{z}) \leq 1$, thus providing a clear quantitative measure of the closeness of a candidate solution to the global optimum.

Figure~\ref{fig:qlstm_qaoa_benchmark} illustrates the loss trajectories over 50 inference iterations for both classical and learning-based optimizers, evaluated on Max-Cut instances with \(P = 3/7\). The results demonstrate that the QLSTM-QAOA framework consistently achieves faster and more stable convergence compared to classical baselines and standard LSTM. Table~\ref{tab:approx_ratio_iter2} reports the approximation ratios (mean~$\pm$~std) at iteration~2 for various values of \(P = 2/7, 3/7, \dots, 6/7\), where \(P\) denotes the \emph{connectivity probability} of each edge in an \textsc{Erd\H{o}s--R\'enyi} random graph. We generated twenty such random instances for each pair \((P, N)\), where \(N \in \{8,\dots,20\}\). Although LSTM and QLSTM optimizers were initially \emph{trained} on 7-node graphs, they maintain strong generalization when tested on larger graphs, converging near their respective minima by \emph{iteration step~3}. We therefore compare the third-iteration step performance of the quantum-enhanced optimizers to standard baselines across all instances. Notably, at lower connectivity (e.g., \(P = 2/7\)), QLSTM exhibits the highest approximation ratio, underscoring its resilience and swift convergence in relatively sparse scenarios. As \(P\) increases toward unity (denser graphs), LSTM and QLSTM both achieve robust approximation ratios (e.g., \(0.92 \pm 0.03\) for QLSTM at \(P = 6/7\)), indicating their suitability for deeper circuit layers and large-scale problem sizes typical of challenging quantum optimization tasks.

\begin{figure}[!t]
    \centering
    \includegraphics[width=\columnwidth]{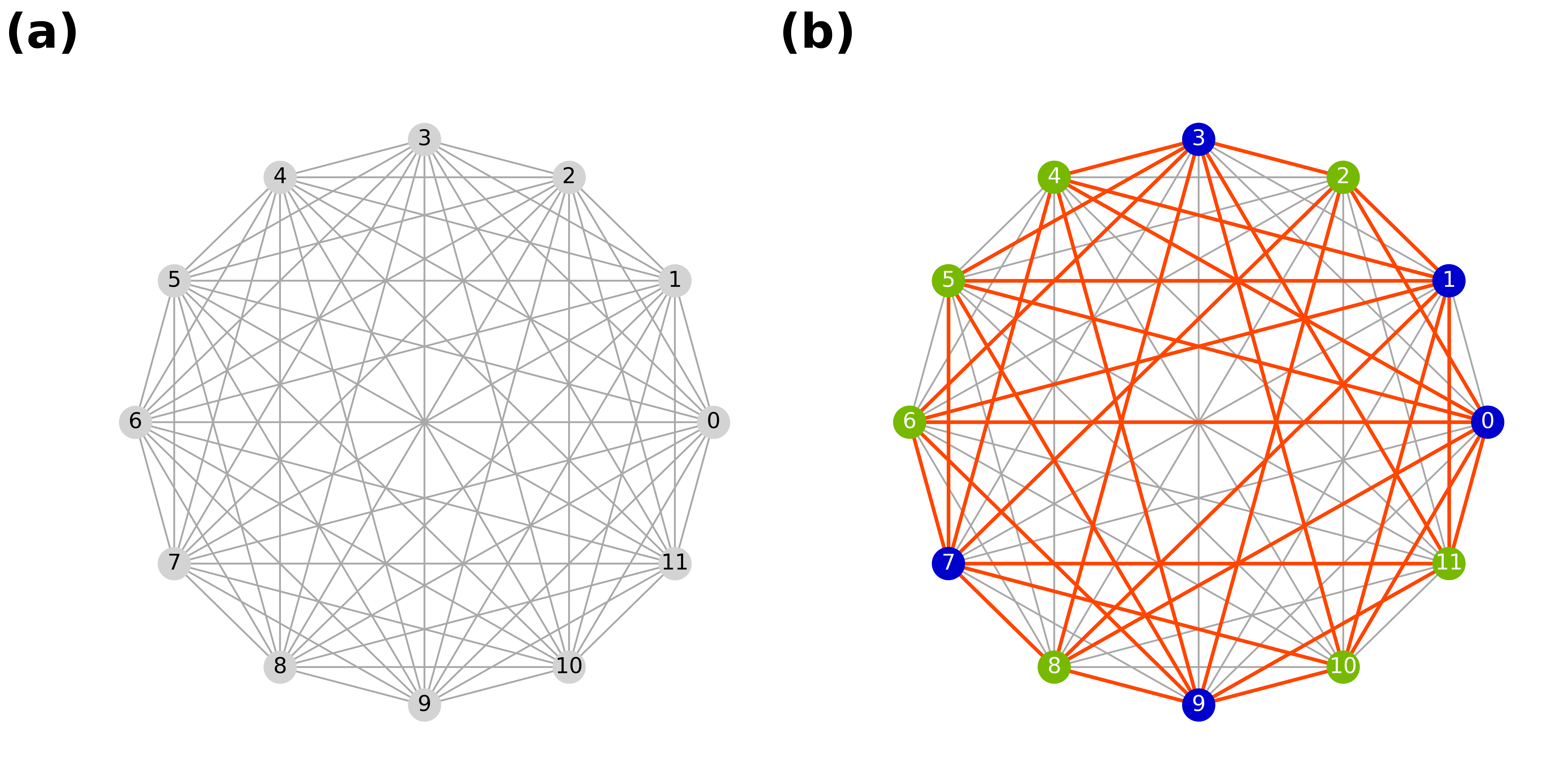}
    \caption{SK Model Problem and a Representative Solution on a 12-Node Fully Connected Graph. (a) Depiction of the SK model instance, illustrating complete connectivity among all $N=12$ spins. (b) An example two-partition assignment, highlighting the cut edges in orangy red.}
    \label{fig:sk_model_architecture}
\end{figure}


\subsection{The Sherrington-Kirkpatrick Model}
\label{subsec:sk_model}

\begin{table*}[!t]
\centering
\caption{Approximation Ratio Table at iteration=3 for \(p=1\), rounded to two decimals.}
\label{tab:approx_ratio_p1_iter3}
\begin{tabular}{ccccccc|c}
\toprule
num\_node & QLSTM & LSTM & RMSProp & SGD & Adam & Adagrad & R-QAOA \cite{finvzgar2024quantum}\\
\midrule
8  & \textbf{1.00} & \textbf{1.00} & 0.64 & 0.58 & 0.42 & 0.42 & 0.92 \\
9  & 0.99 & \textbf{1.00} & 0.66 & 0.73 & 0.45 & 0.45 & 0.89 \\
10 & \textbf{0.99} & \textbf{0.99} & 0.69 & 0.86 & 0.47 & 0.47 & 0.93 \\
11 & \textbf{0.99} & \textbf{0.99} & 0.70 & 0.92 & 0.50 & 0.49 & 0.87 \\
12 & \textbf{0.98} & \textbf{0.98} & 0.72 & 0.93 & 0.51 & 0.51 & 0.92 \\
13 & \textbf{0.98} & \textbf{0.98} & 0.73 & 0.92 & 0.53 & 0.53 & 0.86 \\
14 & 0.97 & \textbf{0.98} & 0.75 & 0.91 & 0.55 & 0.54 & 0.92 \\
15 & \textbf{0.99} & \textbf{0.99} & 0.76 & 0.93 & 0.57 & 0.57 & 0.83 \\
16 & \textbf{1.00} & 0.99 & 0.77 & 0.96 & 0.58 & 0.58 & 0.84 \\
\bottomrule
\end{tabular}
\end{table*}

Originally introduced to capture the physics of spin glasses~\cite{sherrington1975solvable}, the Sherrington-Kirkpatrick (SK) model (shown in Fig.~\ref{fig:sk_model_architecture}) features $N$ classical Ising spins $\bm{z}=(z_1,\ldots,z_N)$ with $z_i\in\{+1,-1\}$, coupled \emph{all-to-all} by random interactions $J_{jk}$. In its canonical form, each instance is defined by a cost function
\begin{equation}
\label{eq:SK-Cz}
C(\bm{z}) \;=\; \frac{1}{\sqrt{N}}\, \sum_{j<k} J_{jk}\,z_j\,z_k,
\end{equation}
where $J_{jk}$ are drawn i.i.d.\ from a mean-zero distribution (e.g.\ $\pm 1$ with equal probability or a normal distribution). The coupling matrix is fully connected, thereby giving rise to a complex, highly frustrated energy landscape. For completeness, we note that in the zero-temperature limit, the Sherrington--Kirkpatrick model exhibits a ground-state cost that scales as \(\approx -0.763\,N\), as shown by Parisi et al.~\cite{parisi1979infinite}.


Within this framework, as in the earlier discussion of the Max-Cut problem, each spin is promoted to a qubit degree of freedom, while the classical coupling $z_i z_j$ is replaced by the Pauli-$Z_i Z_j$ term in the Hamiltonian. Consequently, the objective is to variationally prepare a low-energy state of the system.
\begin{equation}
    H_{\text{SK}} \;=\; -\sum_{j<k}J_{jk}\,Z_j\,Z_k,
\end{equation}
analogous to the Max-Cut Hamiltonian in Eq.~\eqref{eq:HC}, though fully connected and drawn from a random distribution. As we demonstrate below, advanced learning-based optimizers can efficiently navigate this rugged energy landscape, often outperforming conventional approaches by converging to high-quality minima with fewer iterations in practice.

Table~\ref{tab:approx_ratio_p1_iter3} presents the approximation ratios at iteration 3 for \(p=1\) across various optimizers applied to the SK model. Notably, both QLSTM and LSTM consistently yield ratios near unity across system sizes ranging from 8 to 16 nodes, indicating that these learning-based approaches achieve solutions that are nearly optimal relative to the best-known configuration. In contrast, classical optimizers such as RMSProp, SGD, Adam, and Adagrad exhibit significantly lower ratios, with Adam and Adagrad starting around 0.42 for 8 nodes and only reaching approximately 0.58 by 16 nodes. This disparity suggests that the advanced, quantum-inspired strategies are more adept at navigating the complex, highly frustrated energy landscape intrinsic to fully connected spin glasses. The results imply that the superior performance of QLSTM and LSTM may stem from their capacity to capture long-range dependencies and effectively escape local minima, thereby offering a promising pathway for efficiently approximating ground-state configurations in large-scale disordered systems.


\section{Conclusion and Future Work}
\label{sec:cfw}

In this paper, we proposed a hybrid meta-learning approach integrating QLSTM networks as a parameter optimizer for the QAOA. This framework empowers quantum algorithms to ``learn to learn'' across multiple problem instances, enabling swift adaptation and improved solution quality. Our numerical experiments on both the Max-Cut problem and the SK model highlight the QLSTM optimizer’s ability to converge rapidly to near-optimal solutions, outperforming classical gradient-based baselines in terms of both iteration count and final approximation ratios. Further, the QLSTM optimizer’s transfer-learning capability allows it to be trained on smaller graphs and seamlessly deployed on larger instances, yielding significant reductions in runtime and computational overhead.

Looking ahead, an important direction is to extend this meta-learning paradigm to other prominent NISQ-compatible algorithms and to systematically analyze the interplay between circuit depth, hardware noise, and optimizer architecture. Investigating the QLSTM’s internal representations may also offer deeper insights into how quantum-enhanced recurrent networks capture correlations in nontrivial problem landscapes. Moreover, exploring adaptive learning rates or integrating advanced regularization strategies could further bolster robustness against quantum measurement fluctuations. We envision that these future studies will refine the quantum meta-learning paradigm, accelerating the practical deployment of quantum optimization algorithms in realms such as materials science, logistics, and finance.

\section*{Acknowledgment}
This work was performed for the Council for Science, Technology and Innovation (CSTI), Cross-ministerial Strategic Innovation Promotion Program (SIP), ``Promoting the application of advanced quantum technology platforms to social issues" (Funding agency: QST). This work was also supported by the Engineering and Physical Sciences Research Council (EPSRC) under grant number EP/W032643/1.



\bibliographystyle{IEEEtran}
\bibliography{bib/tools,bib/vqc,bib/qml_examples,bib/quantum_fl, bib/ml_examples, bib/hybrid_co_examples,bib/classical_fl,references,bib/fwp,bib/qt}

\end{document}